\documentclass[12pt]{article}
\usepackage[dvips]{graphicx}
\usepackage{amsmath}
\usepackage{amssymb}
\newtheorem{pos}{Postulate}

\title{Borromean Entanglement Revisited}
\author{
Ayumu Sugita
\thanks{sugita@a-phys.eng.osaka-cu.ac.jp}
\\
Osaka City University\\
Sugimoto, Sumiyoshi-ku, Osaka 558-8585, Japan}
\date{}

\begin{document}
\maketitle

\begin{abstract}
An interesting analogy between quantum entangled states and topological
links was suggested by Aravind. 
In particular, he emphasized a connection
between the Greenberger-Horne-Zeilinger (GHZ)
state and the Borromean rings.
However, he made the connection in a way that depends
on the choice of measurement basis.
We reconsider it in a basis-independent way
by using the reduced density matrix.

\end{abstract}

\section{Introduction}

Quantum entanglement is a form of correlation among quantum systems
which cannot be described
by any classical (local reality) theory. 
This strange nonlocality in quantum mechanics, which Albert Einstein called
"spooky action at a distance", is now considered to be
the key resource for quantum information processing 
such as quantum computation and quantum cryptography.

It seems quite natural to think of analogies
between quantum entanglement and topological entanglement, and
some authors actually suggested such analogies
\cite{aravind, kauffman}. Among them,
we pick up the idea of Aravind \cite{aravind}
in this paper. 
He made the correspondence between quantum states and topological links
by associating each particle with a ring, and measurement of 
a particle with cutting of the corresponding ring.
However, there are many possible measurements
of a particle, and the correspondence depends on the
choice of the measurement. We reconsider it
by using the partial trace of the density matrix
instead of the measurement.

In the following, we first review the basics
of quantum mechanics for convenience of the
readers who are not familiar with quantum mechanics and entanglement. 
Then we proceed to
the discussion on the correspondence
between quantum states and links.

\section{Basic concepts and tools}

In this section we introduce basic concepts and 
tools of quantum mechanics according to Chapter 3
of the text book by Nielsen and Chuang \cite{nielsen_chuang}.

\subsection{Postulates of quantum mechanics}

The mathematical structure of quantum mechanics can be 
summarized in the following four postulates.

\begin{pos}
Associated to any isolated physical system is a complex
vector space with inner product (that is a Hilbert space) known as the 
{\it state space} of the system. The system is completely described by
its {\it state vector}, which is a unit vector in the system's state space.
\end{pos}
We use the standard quantum mechanical notation $|\psi \rangle$ to represent a vector,
where $\psi$ is a label for the vector. Its dual vector is denoted as $\langle \psi |$.
$\langle \phi |\psi\rangle$ represents the inner product between the vectors
$|\phi\rangle$ and $|\psi\rangle$. Note that the product of the form
$|\phi\rangle \langle \psi |$ is regarded as an operator which acts
on a vector $|v\rangle$ as $|\phi\rangle \langle \psi |v\rangle$.

\begin{pos}
The evolution of a closed quantum system is described
by a {\it unitary transformation}. 
\end{pos}

\begin{pos}
Ideal quantum measurements are described by a collection $\{P_m\}$ of 
{\it projection
operators}  acting on the state space of the system. 
The index $m$ refers to the measurement outcomes.
If the state of the quantum system is $|\psi\rangle$ immediately before
the measurement then the probability that result
$m$ occurs is 
\begin{eqnarray}
p(m) = \langle \psi |P_m |\psi\rangle, 
\end{eqnarray}
and the state of the system after the measurement is
\begin{eqnarray}
\frac{P_m |\psi\rangle}{\sqrt{\langle \psi|P_m |\psi\rangle}}.
\end{eqnarray}
The projection operators satisfy the relation
\begin{eqnarray}
\sum_m P_m = I,
\end{eqnarray}
where $I$ is the identity operator, so that the probabilities sum to one.
\end{pos}
Postulates 3 describes only ideal (projective) measurements, 
but actual measurements
are rarely ideal. Therefore some text books including 
\cite{nielsen_chuang}
adopt more general description of measurements from the beginning.
However, general measurements can be represented by combining projective
measurements and other postulates of quantum mechanics. 
Since we use only projective measurements in this paper,
we do not include general measurements in Postulate 3 for simplicity.

\begin{pos}
The state space of a composite physical system is the
tensor product of the state spaces of the component physical systems.
Moreover, if we have systems numbered $1$ through $n$, and system number
$i$ is prepared in the state $|\psi_i\rangle$, then the joint state of the
total system is $|\psi_1\rangle \otimes |\psi_2\rangle \otimes \dots \otimes |\psi_n\rangle$. 
\end{pos}
We often use abbreviated notations $|\psi_1\rangle |\psi_2\rangle $ or $|\psi_1 \psi_2\rangle$
to denote $|\psi_1\rangle \otimes |\psi_2\rangle$.

In the following, we consider composite systems consisting of {\it qubits}.
A qubit (quantum bit) is the simplest quantum mechanical system
whose state space is two-dimensional. 
It is used as a unit of quantum 
information. We denote the basis vectors of the two-dimensional Hilbert space as
$|0\rangle$ and $|1\rangle$.

\subsection{The density operator}

The state vector formalism introduced in the last subsection
can be used to describe a system whose state is completely specified.
In such a case, the state is called a {\it pure state}.
In most situations, however, we do not have complete knowledge
about the state. The density operator can be used to describe
such a situation.  


Suppose a quantum system is in one of a
number of states $|\psi_i\rangle$, where $i$ is an index, with respective probabilities $p_i$.
The set $\{p_i, |\psi_i\rangle\}$ is called an ensemble of pure states. 
The density operator for the system is defined as
\begin{eqnarray}
\rho = \sum_i p_i |\psi_i\rangle \langle \psi_i|.
\label{decomp}
\end{eqnarray} 
If $p_i=1$ for an index $i$ and all other probabilities are zero,
the density matrix
\begin{eqnarray}
\rho = |\psi_i\rangle \langle \psi_i|
\end{eqnarray}
describes a pure state corresponding to the state vector
$|\psi_i\rangle$.
Otherwise
the density operator $\rho$ represents a {\it mixed state}.

In the density operator formalism, time evolution of a quantum system is 
described by a unitary operator $U$ as $\rho \rightarrow U\rho U^\dagger$.
The probability that the result labeled by $m$ occurs is
$
{\rm tr} (\rho P_m)
$,
and the the state of the system after the measurement is 
$\frac{P_m \rho P_m}{\sqrt{{\rm tr} (\rho P_m)}}$.
It is easy to check that the density matrix formalism
reduces to the state vector formalism 
when it represents a pure state.

An operator can be regarded as a density operator if and only if 
it is a positive operator with ${\rm tr} \rho = 1$. If an operator satisfies
this condition, it is obvious that
the operator can be written in the form (\ref{decomp}),
where $p_i$ is an eigenvalues of $\rho$ and 
$|\psi_i\rangle$ is the corresponding eigenvector. Therefore
$\rho$ is the density operator associated to the ensemble 
$\{p_i, |\psi_i\rangle \}$. Note, however, that the correspondence
between density operators and ensembles is not one-to-one.
There are infinitely many ensembles corresponding to a density operator.

\subsection{The reduced density operator}

When we observe only a subsystem of a composite system,
the subsystem is described by the reduced density operator.
Suppose we have a physical system composed of 
subsystems A and B, and its state is
described by a density operator $\rho^{AB}$.
The reduced density operator for system A is defined by
\begin{eqnarray}
\rho^{A} \equiv {\rm tr}_B \left(\rho^{AB}\right),
\end{eqnarray}
where ${\rm tr}_B$ is the {\it partial trace} over system B.
The partial trace is defined by
\begin{eqnarray}
{\rm tr}_B \left(|a_1\rangle \langle a_2| \otimes |b_1\rangle\langle b_2|\right)
\equiv
|a_1\rangle \langle a_2|\, {\rm tr} \left(|b_1\rangle \langle b_2|\right),
\end{eqnarray}
where $|a_1\rangle$ and $|a_2\rangle$ ($|b_1\rangle$ and $|b_2\rangle$) 
are any two vectors in the state space
of A (B). In addition, we require the linearity of the partial trace,
which completes the specification.

\subsection{Entanglement}
Let us consider a pure state of a system composed of subsystems A and B.
If the state vector of the system can be written as a tensor product
\begin{eqnarray}
|\psi\rangle = |\psi_A\rangle \otimes |\psi_B\rangle,
\end{eqnarray}
where $|\psi_A\rangle$ ($|\psi_B\rangle$) is some state vector of subsystem A (B),
the state is called {\it separable}. It is known that if a pure state is not
separable, there exists a Bell-type inequality which is violated by this state
\cite{peres_txt}.
It means that the state has a correlation which cannot be described
by any classical (local reality) theory. Such a state
is called {\it entangled}. For example, it is not difficult to
show that the following 2-qubit state
\begin{eqnarray}
\frac{1}{\sqrt{2}}(|00\rangle + |11\rangle) 
= 
\frac{1}{\sqrt{2}}(|0\rangle \otimes |0\rangle \;+\; |1\rangle \otimes |1\rangle) 
\end{eqnarray}
cannot be written as a tensor product of two single-qubit states. Hence
this state is entangled.

An mixed state is called separable if it can be decomposed into
the following form
\begin{eqnarray}
\rho = \sum_i p_i \rho_i^A \otimes \rho_i^B,
\label{sep_def}
\end{eqnarray}
where $\rho^A_i$ ($\rho^B_i$) is a density matrix of subsystem A (B),
$0\le p_i \le 1$ and $\sum_i p_i =1$.
Since $\rho^A_i$and $\rho^B_i$ can be represented by ensembles
of pure states, (\ref{sep_def}) means that
a separable mixed state can be represented by an ensemble of separable
pure states.

In this paper, a non-separable mixed state is called entangled.
Note, however, that the definition of entanglement is
not so trivial for mixed states. For example,
a separable mixed state satisfies all Bell-type inequalities, 
but the converse does not hold \cite{werner}.

In general, it is a difficult task to determine if
a given mixed state is separable or not. However,
there are some simple criteria for 2-qubit systems.
One is Peres' criterion based on the positivity of
the partial transposed density matrix \cite{peres}.
Another useful criterion is based on the {\it concurrence}
\cite{wootters},
which is given by
\begin{eqnarray}
C(\rho ) = 
{\rm max} 
\{ 0,\;\; \lambda_1 - \lambda_2 - \lambda_3 - \lambda_4 \}.
\end{eqnarray}
Here, $\lambda_i$s are the square roots of the eigenvalues of 
$\tilde{\rho}\rho$
in decreasing order, 
where 
\begin{eqnarray}
\tilde{\rho} = (\sigma_y \otimes \sigma_y) \rho^* (\sigma_y \otimes \sigma_y)
\end{eqnarray}
and
\begin{eqnarray}
\sigma_y = 
\left(
\begin{array}{cc}
0 & -i \\
i & 0
\end{array}
\right).
\end{eqnarray}
A state represented by the density operator $\rho$ is separable
if and only if $C(\rho) = 0$.


\section{Quantum and topological entanglement}

\begin{figure}
\begin{center}
\begin{tabular}{ccc}
\begin{minipage}{0.4\hsize}
\begin{center}
\includegraphics[width=1.0\textwidth]{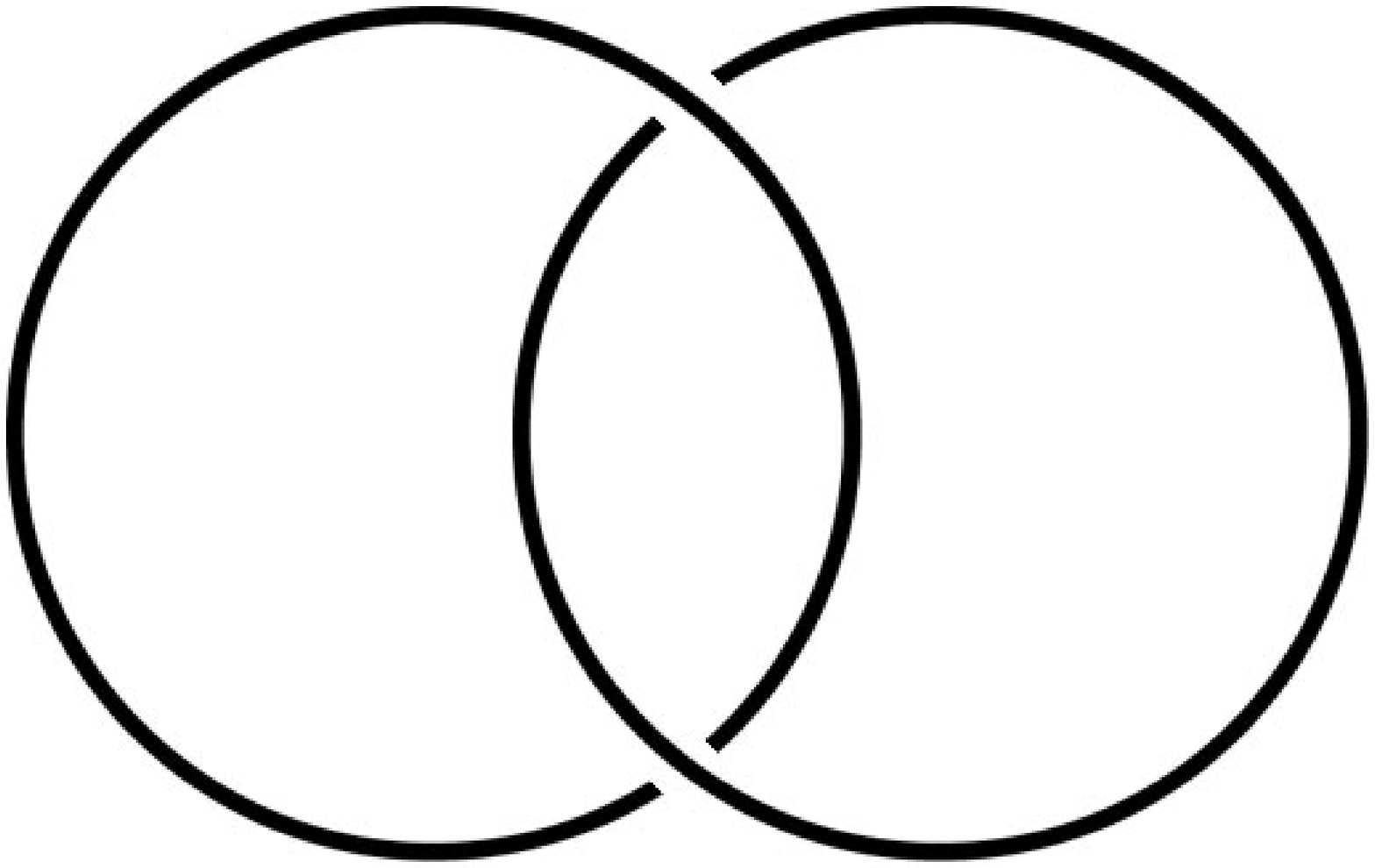}
\caption{Entangled two rings corresponding to entangled two qubits.}
\label{2q_ent}
\end{center}
\end{minipage}
&
\begin{minipage}{0.1\hsize}
\end{minipage}
&
\begin{minipage}{0.4\hsize}
\begin{center}
\includegraphics[width=1.0\textwidth]{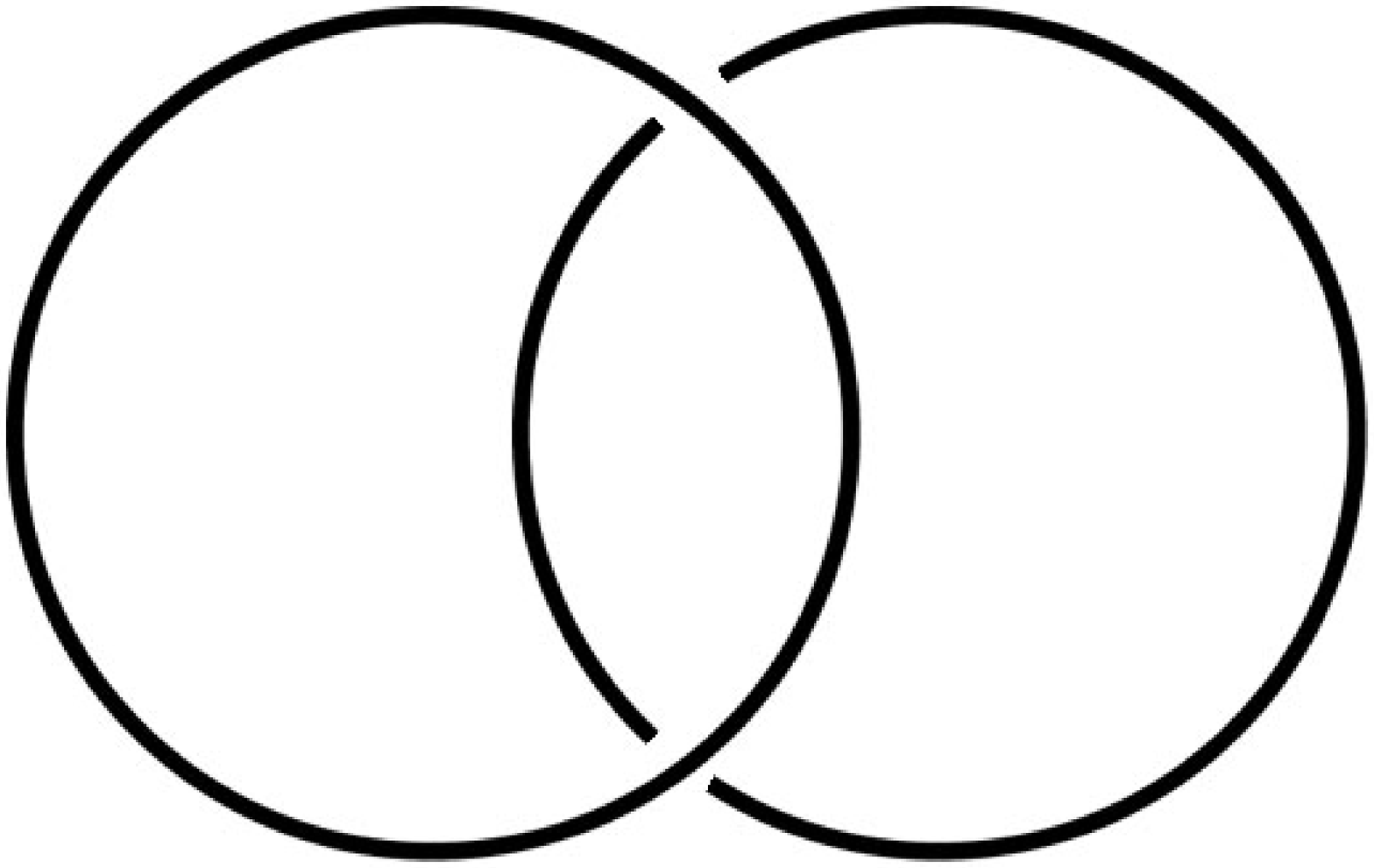}
\caption{Unentangled two rings corresponding to unentangled two qubits.}
\label{2q_unent}
\end{center}
\end{minipage}
\end{tabular}
\end{center}
\end{figure}

Let us introduce a correspondence between quantum states and 
and links. 
In the following, we consider composite systems consisting of qubits,
and associate a ring with a qubit.
Entangled two qubits are represented by entangled two rings
(Fig. \ref{2q_ent}), and separable two qubits are
represented by unentangled two rings (Fig. \ref{2q_unent}).

Then let us consider 3-qubit systems.
The three qubits are named A, B and C,
and the tensor product of the states
of three qubits $|\psi_A\rangle\otimes |\psi_B\rangle \otimes |\psi_C\rangle$ 
is denoted as $|\psi_A \psi_B \psi_C\rangle$,
where $\psi_A$, $\psi_B$ and $\psi_C$ are labels for 
the states of A, B and C, respectively.

As the first example, we consider the Greenberger-Horne-Zeilinger
(GHZ) state
\begin{eqnarray}
|{\rm GHZ}\rangle = \frac{1}{\sqrt{2}}(|000\rangle + |111\rangle).
\end{eqnarray}
This state is entangled, hence the corresponding three rings
are also tangled somehow.

\begin{figure}
\begin{center}
\includegraphics[width=0.50\textwidth]{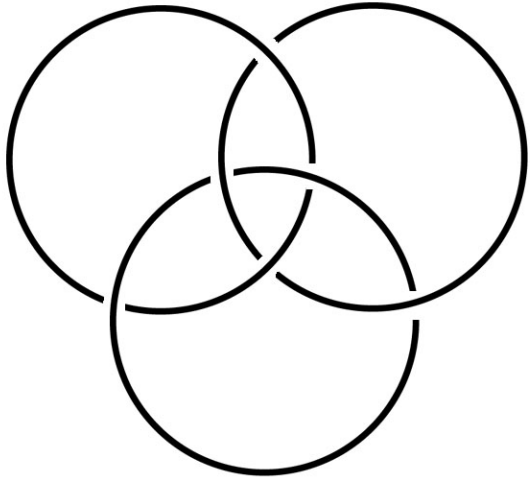}
\caption{The Borromean rings. If one of the three rings is cut, 
the other two rings can be pulled apart.}
\label{borromean}
\end{center}
\end{figure}

Aravind considered a measurement of a qubit whose measurement
operators are $P_0 \equiv |0\rangle\langle 0|$ and 
$P_1 \equiv |1\rangle \langle 1|$
(or more precisely, $P_0 = |0\rangle\langle 0|\otimes I\otimes I$
and $P_1 = |1\rangle\langle 1|\otimes I\otimes I$,
where $I$ is the identity operator for a qubit).
Since the GHZ state is symmetric
under permutations of the three qubits, it is enough to
consider the measurement of a qubit, say, A. 
Then the state of the system after the measurement
is $|000\rangle$ or $|111\rangle$, both of which are completely separable.
If we associate the measurement with cutting of the corresponding ring,
it means that the three rings are separated by cutting only 
one of the three rings. It is very nature of the famous
Borromean rings (Fig. \ref{borromean}).

However, there are many possible measurements for a qubit.
For example, let us consider a measurement corresponding
to the spin measurement along $x$-direction.
The measurement operators are $P_{+} \equiv |+\rangle\langle +|$
and $P_{-} \equiv |-\rangle \langle -|$, where
\begin{eqnarray}
|+\rangle &\equiv& \frac{1}{\sqrt{2}}(|0\rangle + |1\rangle),\\
|-\rangle &\equiv& \frac{1}{\sqrt{2}}(|0\rangle - |1\rangle).
\end{eqnarray}
The GHZ state can be rewritten as
\begin{eqnarray}
|{\rm GHZ}\rangle 
&=&
\frac{1}{2}
\left\{
  |+00\rangle \;+\; |+11\rangle
\;+\; |-00\rangle \;-\; |-11\rangle 
\right\}.
\end{eqnarray}
Therefore the state after the measurement
is 
\begin{eqnarray}
|+00\rangle \;+\; |+11\rangle
&=& |+\rangle \otimes (|00\rangle + |11\rangle)
\end{eqnarray} 
or
\begin{eqnarray} 
|-00\rangle \;-\; |-11\rangle
&=&
|-\rangle \otimes (|00\rangle - |11\rangle),
\end{eqnarray}
 both of which
have entanglement between qubit B and qubit C.
Thus the correspondence introduced by Aravind depends on
the choice of measurement basis.

We use the partial trace instead of measurement
as a counterpart of the cutting of a ring.
Physically speaking, it means that we just ignore a qubit, and 
observe only the other two. 
If we "trace out" qubit A in the GHZ state, the density operator
for the remaining system is
\begin{eqnarray}
\rho^{BC} \equiv {\rm tr}_A \rho =  \frac{1}{2}\left(|00\rangle \langle 00| 
+ |11\rangle \langle 11|\right).
\end{eqnarray}
This is a separable mixed state, because it is 
represented by an ensemble consisting of 
separable pure states 
$|00\rangle$ and $|11\rangle$. 
\footnote{Aravind described
this state as an "entangled" state,
and associated it with two rings which cannot be pulled apart.
It is normal, however, to classify it as an unentangled state.}
Therefore this state corresponds to the Borromean rings.
Thus we can establish a connection between qubits and rings
in a basis-independent way.

\begin{figure}
\begin{center}
\includegraphics[width=0.50\textwidth]{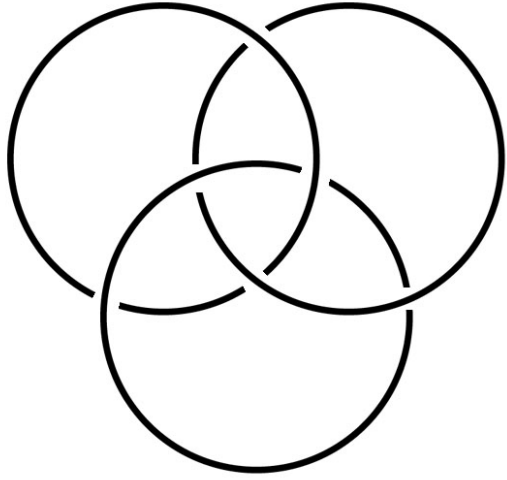}
\caption{Three rings corresponding to the W state.
If any ring is cut, the other two rings are still linked.}
\label{W}
\end{center}
\end{figure}

The next example is the so-called W state
\begin{eqnarray}
|{\rm W}\rangle \equiv \frac{1}{\sqrt{3}}
(|001\rangle + |010\rangle + |100\rangle).
\end{eqnarray}
This state also has the permutation symmetry.
If we trace out qubit A, the reduced density matrix is
\begin{eqnarray}
\rho^{BC} = 
\frac{1}{3}\left(|00\rangle \langle 00| +
|01\rangle\langle 01| + |01\rangle \langle 10|
+ |10\rangle\langle01| + |10\rangle \langle 10| \right),
\end{eqnarray}
which is not separable. Actually we can show the 
non-separability by explicit evaluation of the concurrence
as follows.
The density matrix $\rho^{BC}$
and its conjugate $\tilde{\rho}^{BC}$ can be written in the matrix form as
\begin{eqnarray}
\rho^{BC}
=
\left(
\begin{array}{cccc}
1/3 & 0 & 0 & 0 \\
0 & 1/3 & 1/3 & 0 \\
0 & 1/3 & 1/3 & 0 \\
0 & 0 & 0 & 0
\end{array}
\right),
\;\;\;
\tilde{\rho}^{BC}
=
\left(
\begin{array}{cccc}
0 & 0 & 0 & 0 \\
0 & 1/3 & 1/3 & 0 \\
0 & 1/3 & 1/3 & 0 \\
0 & 0 & 0 & 1/3
\end{array}
\right).
\end{eqnarray}
Hence
\begin{eqnarray}
\tilde{\rho}^{BC}
\rho^{BC}
=
\left(
\begin{array}{cccc}
0 & 0 & 0 & 0 \\
0 & 2/9 & 2/9 & 0 \\
0 & 2/9 & 2/9 & 0 \\
0 & 0 & 0 & 0
\end{array}
\right).
\end{eqnarray}
Therefore
$\lambda_1 = 2/3$, $\lambda_2 = \lambda_3 = \lambda_4 = 0$ and
$C(\rho^{BC})=2/3>0$, which means that, in contrast to the Borromean rings,
each pair of rings cannot be separated even after 
the third ring is cut. This case is modeled by the "three-Hopf rings" in Fig. \ref{W}.

\begin{figure}
\begin{center}
\includegraphics[width=0.60\textwidth]{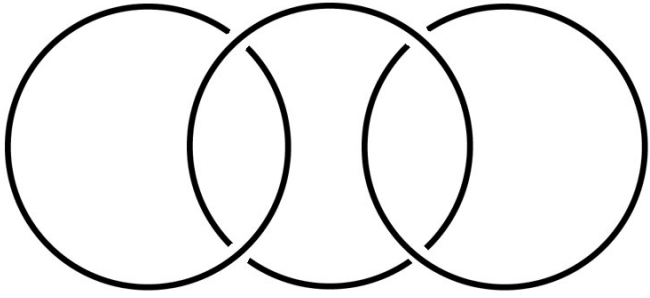}
\caption{Linear-chain configuration of three rings.
If the central ring is cut, the other two rings are unlinked.
But if either edge ring is cut, the other two remain linked.}
\label{chain}
\end{center}
\end{figure}

There is yet another type of 3-qubit entanglement.
Let us consider the following state
\begin{eqnarray}
|\psi\rangle &=& 
a|000\rangle + b|+1+\rangle\\
&=&
a |000\rangle 
+ \frac{b}{2} (|010\rangle + |011\rangle + |110\rangle + |111\rangle).
\label{chain_state}
\end{eqnarray}
with $a, b \in \mathbb{R}$ and $a^2 + b^2 = 1$.
This state 
does not have the full permutation symmetry, but is symmetric
with respect to qubit A and qubit C. If we trace out qubit B, 
the remaining state is
\begin{eqnarray}
\rho^{AC} = a^2 |00\rangle \langle 00|
\;+\; b^2 |++\rangle \langle ++ |,
\end{eqnarray}
which is separable. If we trace out qubit A or C, however,
the remaining state is still entangled. 
For example, if we trace out qubit C, the reduced density operator
$\rho^{AB}$ and its conjugate $\tilde{\rho}^{AB}$
can be written in the matrix form as
\begin{eqnarray}
\rho_{AB} 
&=&
\left(
\begin{array}{cccc}
a^2 & 0 & ab/2 & ab/2 \\
0 & 0 & 0 & 0 \\
ab/2 & 0 & b^2/2 & b^2/2 \\
ab/2 & 0 & b^2/2 & b^2/2
\end{array}
\right),
\;\;\;
\tilde{\rho}_{AB}
=
\left(
\begin{array}{cccc}
b^2/2 & -b^2/2 & 0 & ab/2 \\
-b^2/2 & b^2/2 & 0 & -ab/2 \\
0 & 0 & 0 & 0 \\
ab/2 & -ab/2 & 0 & a^2
\end{array}
\right).
\end{eqnarray}
Hence
\begin{eqnarray}
\tilde{\rho}_{AB} \rho_{AB} 
&=&
\frac{ab}{4}
\left(
\begin{array}{cccc}
3ab & 0 & 2b^2 & 2b^2 \\
-3ab & 0 & -2b^2 & -2b^2 \\
0 & 0 & 0 & 0 \\
4a^2 & 0 & 3ab & 3ab
\end{array}
\right). 
\end{eqnarray}
Then we obtain 
$\lambda_1 = \frac{\sqrt{2} + 1}{2}|ab|$,
$\lambda_2 = \frac{\sqrt{2} - 1}{2}|ab|$,
$\lambda_3 = \lambda_4 = 0$, 
and hence $C(\rho^{AB}) = C(\rho^{BC})=|ab|$, which is positive
if $ab\ne 0$. This result means that the state (\ref{chain_state}) is modeled
by the rings of linear-chain type in Fig. \ref{chain}. 

We have shown some examples of the correspondence between
quantum entanglement and topological entanglement for
some 3-qubit cases. It would be interesting to
construct such correspondences for systems with
more qubits. For example, Aravind argued a
correspondence between the generalized GHZ state
\begin{eqnarray}
\frac{1}{\sqrt{2}}(|00\dots 0\rangle + |11\dots 1\rangle)
\end{eqnarray}
and the generalized Borromean rings.
Although our method does not uniquely 
determine
a topological link corresponding to a quantum state,
visualization of quantum states
by topological objects could be a useful tool for
the study of quantum entanglement.


\begin{thebibliography}{99}

\bibitem{aravind}
P. K. Aravind,
{\it Quantum Potentiality, Entanglement and Passion-at-a-Distance: 
Essays for Abner Shimony}, 
eds. R. S. Cohen, M. Horne and J. Stachel, Kluwer, Dordrecht (1997), pp53-59.

\bibitem{kauffman}
L. H. Kauffman and S. J. Lomonaco, 
New J. Phys. {\bf 4} (2002), pp73.1-73.18.

\bibitem{nielsen_chuang}
M. A. Nielsen and I. L. Chuang,
{\it Quantum Computation and Quantum Information},
Cambridege University Press (2000).

\bibitem{peres_txt}
A. Peres,
{\it Quantum Theory: Concepts and Methods},
Kluwer Academic Publishers.

\bibitem{werner}
R. R. Werner,
Phys. Rev. A {\bf 40} (1989) 4227.

\bibitem{peres}
A. Peres,
Phys. Rev. Lett. {\bf 77} (1996) 1413.


\bibitem{wootters}
W. K. Wootters,
Phys. Rev. Lett. {\bf 80} (1998) 2245.

\end{thebibliography}
\end{document}